\documentclass[hyper,cits]{PoS}
\usepackage{epsfig}
\usepackage{amsmath,amssymb}

\title{Monte Carlo approach to turbulence\footnote{DESY 09-174, MS-TP-09-23}}

\ShortTitle{Monte Carlo approach to turbulence}

\author{P.~D\"uben$^{a}$, D.~Homeier$^{a}$, K.~Jansen$^{b}$, 
D.~Mesterhazy$^{c}$ and \speaker{G.~M\"unster}$^{\, a}$\\
$^{a}$ Westf\"alische Wilhelms-Universit\"at, Intitut f\"ur 
Theoretische Physik\\
Wilhelm-Klemm-Str.~9, 48149 M\"unster, Germany\\
$^{b}$ NIC, DESY Zeuthen\\
Platanenallee 6, 15738 Zeuthen, Germany\\
$^{c}$ Humboldt-Universit\"at zu Berlin, Institut f\"ur Physik\\
Newtonstrasse 15, 12489 Berlin, Germany\\
E-mail: \email{munsteg@uni-muenster.de}}

\abstract{The behavior of the one-dimensional random-force-driven Burgers
equation is investigated in the path integral formalism on a discrete
space-time lattice. We show that by means of Monte Carlo methods one may
evaluate observables, such as structure functions, as ensemble averages over
different field realizations. The regularization of shock solutions to the
zero-viscosity limit (Hopf-equation) eventually leads to constraints on
lattice parameters required for the stability of the simulations. Insight
into the formation of localized structures (shocks) and their dynamics is
obtained.}

\FullConference{The XXVII International Symposium on Lattice Field Theory\\
		 July 26 - 31, 2009\\
		 Peking University, Beijing, China}
		 
\begin{document}

\section{Introduction}
Besides tremendous research having been done since Kolmogorov's famous
publication in 1941 \cite{1}, hydrodynamic turbulence essentially remains an
unsolved problem of modern physics. This is especially remarkable as the
fundamentals seem to be fairly easy -- the Navier-Stokes equations for the
velocity field $u_{\alpha}$ and pressure $p$
\begin{equation}
    \partial_{t}u_{\alpha} + u_{\beta} \partial_{\beta} u_{\alpha} 
    - \nu \nabla^{2} u_{\alpha}
  + \frac{1}{\rho}\partial_{\alpha}p = 0
\end{equation}
with the additional constraint
\begin{equation}  \partial_{\alpha}u_{\alpha} = 0
\end{equation}
simply express the conservation of momentum in a classical, incompressible
fluid of viscosity $\nu$ and density $\rho$. For laminar flows it is well
known that the Navier-Stokes equations reproduce realistic flows very
accurately; in the turbulent regime, it is still an open question how the
universal characteristics of turbulent flow, characterized by the scaling
exponents $\xi_p$ of structure functions $S_p$ of order $p$, defined by
\begin{equation}
S_p(x) := \overline{ | u(r+x) - u(r) |^{p} } \sim |x|^{\xi_p}, 
\label{dh:sp}
\end{equation}
can be extracted from first principles. Here the bar corresponds to a
spatial averaging.

Monte Carlo simulations in the path integral formulation enable us to gain
direct insight into the formation of localized structures and their
behavior, and to measure observables as, e.g.\ structure functions and their
scaling exponents \cite{2,3,4}.

\section{Burgers' Equation}

We decided to elaborate the methods using the stochastically forced Burgers
equation \cite{5} in 1+1 dimensions
\begin{equation}
    \partial_{t}u + u \partial_xu - \nu \partial_x^{2} u = f\,,
\label{dh:eqburg}
\end{equation}
which may be interpreted as the flow equation for a fully compressible fluid.
The stochastic force is modeled to be Gaussian with correlation
\begin{equation}\label{dh:eqchi}
 \chi(x,t; x',t') := \big\langle f(x, t) f(x', t') \big\rangle 
 = \epsilon \, \delta(t-t') \exp\left(-\frac{|x-x'|}{\Lambda}\right),
\end{equation} 
where $\Lambda$ defines the correlation length of the forcing and the
$\langle\, \cdots \,\rangle$ denotes the ensemble average. A finite
viscosity $\nu$ and energy dissipation $\epsilon$ provide a dissipation
length scale $\lambda$ corresponding to the Kolmogorov-scale in
Navier-Stokes turbulence:
\begin{equation}
\lambda := \left( \frac{\nu^{3}}{\epsilon}\right)^{\frac{1}{4}}.
\end{equation}
We can furthermore identify the Reynolds-number as
\begin{equation}
\mathit{Re} := (\epsilon \Lambda^4/\nu^3)^{1/3}.
\end{equation}

The fundamental solutions to the Burgers equation are well-known -- in the
limit of vanishing viscosity (Hopf-equation) these form singular shocks. A
finite dissipation scale $\lambda \sim \nu / U$, where $U$ is the
characteristic velocity, provides an UV-regularization of the shock
structures:
\begin{equation}
u = -U \tanh\frac{U}{2\nu}x\,.
\end{equation}
Most interestingly, the exponents $\xi_p$ as defined in (\ref{dh:sp}) are
non-trivial for Burgers turbulence; for the forcing (\ref{dh:eqchi}) and in
the regime $x \sim \lambda$ we have the analytic result \cite{6}
\begin{equation}
 \xi_p = \text{min} (1, p).
\end{equation}

\section{Path Integral Formulation}

Following the method of Martin, Siggia and Rose \cite{7}, we established a
path integral for Burgers' equation
\begin{equation}
Z \propto \int \mathcal{D}u \, \exp\bigg(-\frac{1}{2}\int dt dx \, 
\big(\partial_{t} u+ u \partial_x u
- \nu \partial_x^{2} u \big) \chi^{-1} \ast 
  \big(\partial_{t} u+ u\partial_x u
- \nu \partial_x^{2} u\big) \bigg),
\end{equation}
where $\ast$ denotes the convolution.

It has been shown by Falkovich et al.
\cite{Falkovich:1995fa,Balkovsky:1997zz} on the basis of an equivalent sum
of states that the fundamental solutions of Burgulence can be understood as
instantons.

\section{Monte Carlo Simulations}

For 1+1 dimensional Burgulence, a large number of stable simulations could
be performed; we are working on the final analysis. Typical lattice sizes
range from $(N_x=16) \times (N_t=16)$ up to $(N_x = 4096) \times (N_t =
128)$ lattice points.
\begin{figure}
\begin{center}
  \vspace{1.5cm}
  \includegraphics[width=.7\textwidth]{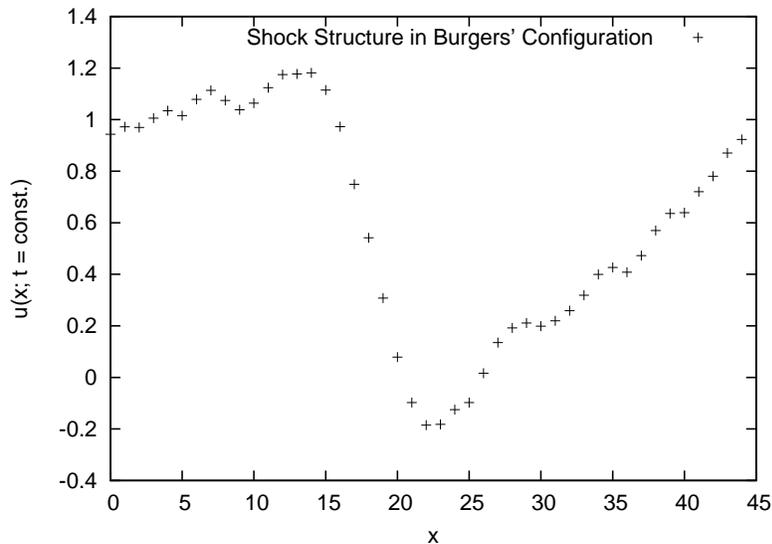}
  \caption{Time-slice of a $(N_x = 256) \times (N_t = 45)$-lattice 
     at $\mathit{Re} = 4$, $\nu = 1/32$, showing the velocity 
     \mbox{$u(x,t = \mathit{const})$} as a function of $x$.
     The typical shock structure is clearly visible.}
  \nobreak\medskip
\end{center}
\end{figure}

\subsection{Boundary Conditions}

To be in general agreement with literature and analytic calculations, we
started with lattices periodic both in time and space direction. In an
attempt to reduce autocorrelation times, we dropped these boundary
conditions. While autocorrelation times did not change much, simulating with
free boundaries effectively doubles the spatial lattice extent and gives
access to excitations of the Burgers vacuum state.

\subsection{Lattice Discretization}

Once having discretized the path integral on a Euclidean lattice of 
spacings $\Delta x$ and $\Delta t$, we get for $\nu$:
\begin{equation} 
\nu = \alpha \frac{(\Delta x)^2}{\Delta t}.
\label{dh:eqnu}
\end{equation}
The continuum limit of our lattice theory is reached by holding $\nu$ and
$\mathit{Re}$ constant while increasing the number of lattice sites.
$\alpha$ is an a priori arbitrary constant that can be interpreted as a
measure for $\Delta t$ depending on $\Delta x$ and also has to be kept fixed
while performing the continuum limit.

\subsection{Algorithms}

We use a local heat bath algorithm with successive over-relaxation (SOR) for
the Monte Carlo evaluation of the partition function \cite{Adler:1981sn}.
The use of certain acceleration techniques with SOR, specifically Chebyshev
acceleration \cite{Varga:1962}, significantly accelerates the thermalization
process.

Though suiting our purposes so far, it poses certain restrictions on
parallelization. We therefore started employing a Hybrid Monte Carlo
algorithm that we expect to scale better with the number of parallel
processors.

\subsection{Autocorrelation Times}

With $\chi$ being a nonlocal operator one would expect long autocorrelation
times in the simulation of stochastically forced differential equations.
However, with the over-relaxed heat bath algorithm and an appropriate
definition of structure functions on the lattice (where the reference point
for evaluation is chosen randomly for each configuration) the integrated
autocorrelation time is reduced to $\tau \sim O(1)$.

\subsection{Resources}

For testing purposes small lattices may easily be simulated on desktop PCs.
However, high resolution simulations on large lattices require massively
parallel architectures. We have run our simulations on the IBM p690 cluster
JUMP at FZ J\"ulich and on the Linux cluster at Humboldt University Berlin
with up to 256 processors in parallel. In July 2009 we continued our
simulations on the new supercomputer JUROPA at FZ J\"ulich.

\section{First Results}

First results include further constraints that have to be imposed in order
to ensure stable numerics. Most constricting is the need to resolve the
Kolmogorov-length scale $\lambda$ on the lattice. We can in this way show
the effect of $\lambda$ as UV-regularization of the otherwise singular
shocks. This translates into a relation for the Reynolds-number:

\begin{equation}
 \mathit{Re} < \frac{\Lambda}{\Delta x} .
\end{equation}

This will become crucial for Navier-Stokes turbulence enforcing us to
simulate big lattices.

\subsection{Structure Functions}

From analytic calculations \cite{6} we have
\begin{equation}
S_p(x) \sim C_p |x|^{p} + C_{p}' |x|,
\end{equation}
for small seperations in the inertial range.

Though our results are in general agreement with this, the extraction of
scaling exponents is far from trivial and very sensitive to statistical
errors.

\begin{figure}
\begin{center}
  \vspace{1.5cm}
  \includegraphics[width=.7\textwidth]{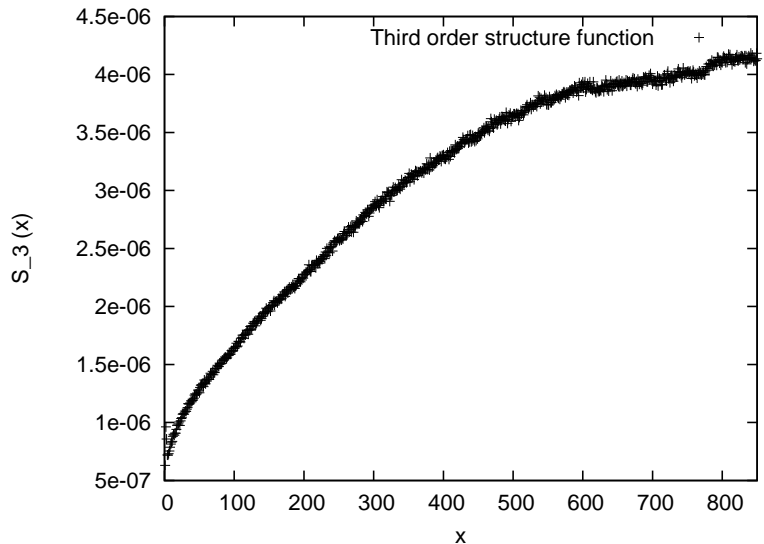}
  \caption{Third order structure function $S_3(x)$ as a function of 
           space separation $x$.}
  \nobreak\medskip
\end{center}
\end{figure}

\subsection{Extended Self-Similarity (ESS)}

Rather than measuring the scaling exponents $\xi_p$ directly, there have
been attempts to measure the scaling behavior of ratios of structure
functions \cite{9}. It was shown that this greatly enhances the inertial
range not only at high but also moderate Reynolds numbers. However, we must
stress that up to now it is not clear if there are any systematic effects in
the evaluation of the structure function exponents via ESS.

\begin{figure}
\begin{center}
  \vspace{1.5cm}
  \includegraphics[width=.7\textwidth]{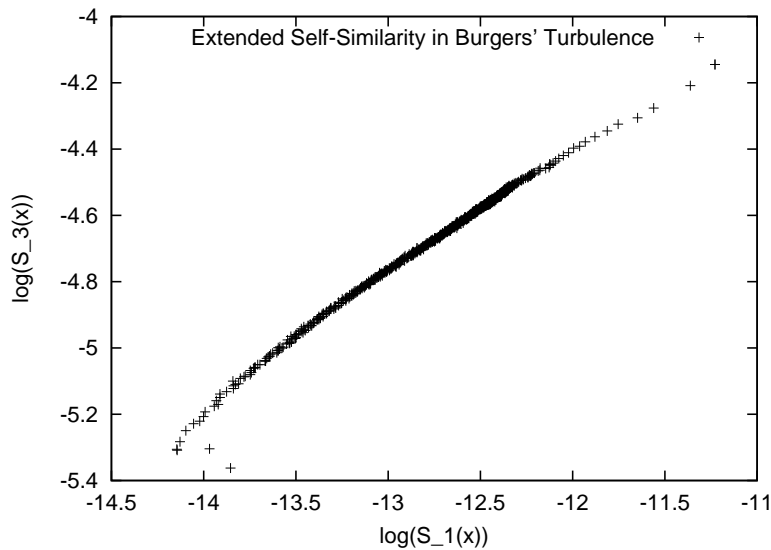}
  \caption{$\log[S_3(x)]$ as function of $\log[S_1(x)]$ clearly showing the 
           linear ESS-dependence.}
  \nobreak\medskip
\end{center}
\end{figure}

\subsection{Outlook}

After completing the analysis of 1+1 dimensional Burgulence, we will proceed
to 3+1 dimensions. The ultimate challenge will be the simulation and
analysis of 2+1 and 3+1 dimensional Navier-Stokes turbulence.

\subsection{Acknowledgements}

We thank the John von Neumann-Institute for Computing (NIC), J\"ulich, for
computing time and support. D.~M. thanks the IRZ Physik at HU Berlin for
computing time spent on the local Linux cluster and their staff for
technical support.

\end{document}